\documentclass[prb, twocolumn,superscriptaddress,preprintnumbers,a4paper,amsmath,amssymb,showpacs,floatfix]{revtex4-1}

\usepackage{graphicx}
\usepackage{color}\color[rgb]{0.000,0.000,0.000} 
\usepackage{amssymb} 
\usepackage{amsmath} 
\usepackage[utf8]{inputenc} 

\usepackage{hyperref}
\usepackage{lineno} 

\begin{document}

\newcommand{\degree}{$^{\circ}$C}

\newcommand{\la}{La$_{5.4}$WO$_{12-\delta}$}

\title{Nanoscale order in the frustrated mixed conductor \la}
\author{Tobias Scherb}
\email[Email of corresponding author:]{tobias.scherb@helmholtz-berlin.de}
\affiliation{Helmholtz-Zentrum Berlin f\"ur Materialien und Energie (HZB), Hahn-Meitner-Platz 1, D-14109, Berlin, Germany;}
\author{Simon A. J. Kimber}
\affiliation{European Synchrotron Radiation Facility (ESRF), 6 rue Jules Horowitz, BP 220, 38043  Grenoble Cedex 9, France;}
\author{Christiane Stephan}
\affiliation{Helmholtz-Zentrum Berlin f\"ur Materialien und Energie (HZB), Hahn-Meitner-Platz 1, D-14109, Berlin, Germany;}
\author{Paul. F. Henry}
\affiliation{European Spallation Source ESS AB, Box 176, 22100, Lund, Sweden;}
\author{Gerhard Schumacher}
\affiliation{Helmholtz-Zentrum Berlin f\"ur Materialien und Energie (HZB), Hahn-Meitner-Platz 1, D-14109, Berlin, Germany;}
\author{Justus Just}
\affiliation{Helmholtz-Zentrum Berlin f\"ur Materialien und Energie (HZB), Hahn-Meitner-Platz 1, D-14109, Berlin, Germany;}
\author{Sonia Escol\'astico}
\affiliation{Instituto de Tecnolog\'ia Qu\'imica (UPV-CSIC)
av. los Naranjos s/n, 46022 Valencia, Spain;}
\author{Jos\'e M. Serra}
\affiliation{Instituto de Tecnolog\'ia Qu\'imica (UPV-CSIC)
av. los Naranjos s/n, 46022 Valencia, Spain;}
\author{Janka Seeger}
\affiliation{Forschungszentrum J\"ulich GmbH (FZJ), 
52425 J\"ulich, Germany.}
\author{Adrian H. Hill}
\altaffiliation[Currently at:]{Johnson Matthey Technology Centre, Intercat-JM Technology Centre, 107 Eli Whitney Boulevard,
Savannah, GA 31408, USA.
.}
\affiliation{European Synchrotron Radiation Facility (ESRF), 6 rue Jules Horowitz, BP 220, 38043  Grenoble Cedex 9, France;}
\author{John Banhart}
\affiliation{Helmholtz-Zentrum Berlin f\"ur Materialien und Energie (HZB), Hahn-Meitner-Platz 1, D-14109, Berlin, Germany;}
\date{\today}

\begin{abstract}
We report a comprehensive investigation of the average and local structure of  \la, which has excellent mixed proton, electron and oxide ion conduction suitable for device applications. Synchrotron X-ray and neutron powder diffraction show that a cubic fluorite supercell describes the average structure, with highly disordered lanthanum and oxide positions. On average the tungsten sites are six-fold coordinated, and we detect a trace (4.4(2) \%) of anti-site disorder. In addition to sharp Bragg reflections, strong diffuse neutron scattering is observed, which hints at short-range order. We consider plausible \textit{local} configurations, and show that the defect chemistry implies a simple 'chemical exchange' interaction that favours ordered WO$_{6}$ octahedra. Our local model is confirmed by synchrotron x-ray pair distribution function analysis and EXAFS experiments performed at the La K and W L$_{3}$-edges. We show that ordered domains of around $\sim$3.5 nm are found, implying that mixed conduction in \la \ is associated with a defective glassy-like anion sublattice. The origins of this ground state are proposed to lie in the non-bipartite nature of the \textit{fcc} lattice and the pairwise interactions which link the orientation of neighbouring octahedral WO$_{6}$ sites. This 'function through frustration' could provide a means of designing new mixed conductors.
\end{abstract}

\maketitle
\section{\label{sec:level1}INTRODUCTION}
Mixed conductors- which transport charge by the motion of ionic species- are essential ingredients in next-generation energy sources such as fuel cells and for separating hydrogen from mixed gas streams in fossil fuel power plants.\cite{yang,fabbri,norby:2007,jordal}  Such applications have demanding requirements for mixed conductors that can withstand high temperatures and reactive atmospheres. Proton conducting ceramic membranes will also play a vital role in challenging industrial processes such as gas-to-liquid conversion, selective dehydrogenation, water-gas-shift reaction and ammonia synthesis thanks to their integration in catalytic membranes reactors.\cite{jordal,fontaine,li}
The crystal structure of mixed conductors plays a crucial role. A combination of a crystalline sublattice, to provide rigidity, and a defective, short-range correlated sublattice of charge carriers would be ideal.\cite{kreuer} Oxygen vacancies in particular are needed for proton conduction, and are usually created by doping. However, this synthetic strategy faces a delicate balancing act\cite{fabbri,arico,pergolesi,malavasi} since the creation of defects by doping causes structural inhomogeneities, such as cation size and charge mismatches or grain boundaries, which may trap charge carriers.\cite{hempelmann} High-quality grain-boundary free thin films are thus needed for optimum performance.\cite{pergolesi} Proton conduction is also favored by high symmetry lattices. However, in the commonly studied ABO$_{3}$ perovskite materials, the tolerance factor required for cubic structures presupposes large A-site cations such as barium, which result in chemical instabilities in the acidic atmospheres found in typical applications. This is especially true for high-temperature applications, such as the removal of hydrogen from syngas mixtures, which are needed to support the transition to a hydrogen-based economy and to improve the efficiency of current sustainable energy production.\cite{norby:2007}

Rare earth tungstates with the general formula Ln$_{6}$WO$_{12}$ (Ln = rare earth) are promising materials as they show several ordered crystal structures as a function of rare earth size, and an excellent combination of proton and electron conductivity.\cite{mccarthy,shimura,esc:2009,haugsrud} The ceramic La$_{6-x}$WO$_{12-\delta}$ (0.3\ \textless \ \textit{x}\ \textless \ 0.7) provides the highest proton conductivity and fulfills the requirements for an efficient mixed conductor. The physical properties shown by\ \la \ membranes are ideal for applications.\cite{solis,esc:2013} Despite many attempts,the  polymorphism and extremely subtle structural distortions shown by this family of materials have prevented solution of the crystal structure. Various reports of disordered pyrochlore or ordered defect fluorite-like structures have previously been published.\cite{mccarthy,chang} Most recently, Magraso et al.\cite{mag:2009,mag:2012,mag:2013} have published three different structures for  \la \ using neutron and X-ray powder diffraction. These authors eventually converged on a cubic fluorite-type structure with a 2\textit{a}\ x 2\textit{a}\ x 2\textit{a} superlattice caused by cation ordering, and noted that single-phase specimens were only observed for La/W ratios between 5.3 and 5.7.\cite{mag:2009} DFT calculations and recent high temperature neutron diffraction experiments have updated the former model.\cite{mag:2012,mag:2013} However, the exact relationship between the average and local crystal structure, microstructure and proton conductivity remains unresolved. Most fundamentally, it is also presently unclear why the cubic phases of \textit{RE}$_{6-\delta}$WO$_{12-\delta}$ \ stoichiometry are so disordered, given the presence of completely ordered phases very close by in the phase diagram. For example, orthorhombic phases of stoichiometry \textit{RE}$_{10}$W$_{2}$O$_{21}$ \ and rhombohedral \textit{RE}$_{6}$WO$_{12}$ \ phases form for smaller rare earths. Both of these structure types have ordered arrangements of anion defects and hence are very poor mixed conductors.\cite{scherb,haugsrud,lashtabeg,bevan,diot}

Here, we synthesize pure samples with a composition of \ \la \ that show exceptional device performance and stability.\cite{esc:2013} Complementary X-ray and neutron diffraction techniques are used to solve the average crystal structure, which goes beyond recently proposed models. We establish the structure of \ \la, e.g. lattice symmetry, degree of cation ordering and the location of oxygen vacancies. Real-space pair distribution function analysis of high-energy synchrotron X-ray diffraction data is applied to confirm the model of the average crystal structure and to develop a model of the local crystal structure since total scattering analysis provides a tool to model Bragg and diffuse scattering simultaneously.\cite{shoemaker} X-ray absorption spectroscopy (EXAFS) at La K-edge and W L$_3$-edge was performed to confirm the local crystal structure.

\section{\label{sec:level1}EXPERIMENTAL}
\noindent\textit{Sample synthesis}: The powder material was prepared using the corresponding anhydrous oxides as precursors. The reagents were milled in ethanol for 24 hours, dried and calcined at 800\degree \ and 1000\degree \ for 15 hours respectively. Subsequently the mixture was milled again, pressed and then sintered at T\ =\ 1400\degree. A stoichiometry of \la \ was selected to achieve phase-pure materials since nominal La$_{6}$WO$_{12}$ has been shown to segregate La$_{2}$O$_{3}$.\\
\textit{Compositional analysis}: Chemical composition was studied by neutron activation analysis (NAA) at the BERII reactor in Berlin and by inductively coupled plasma optical emission spectrometry (ICP-OES). \\
\textit{Phase and structural analysis}: Phase analysis was performed applying X-ray diffraction using a Bruker D8 diffractometer in Bragg-Brentano geometry and Cu-K$\alpha$ radiation. The crystal structure was studied in detail by different diffraction methods. High-resolution synchrotron X-ray diffraction (HRSXRD) was performed at ESRF beamline ID31 ($\lambda$\ =\ 0.3962\ \AA, T\ =\ 10\ K and 295\ K). Neutron powder diffraction (ND) data were collected on the D1A and D2B diffractometers at the Institut Laue-Langevin, Grenoble, at wavelengths of $\lambda$\ =\ 1.909\ \AA \ (T\ =\ 573\ K) and $\lambda$\ =\ 1.595 \AA \ (T\ =\ 300\ K), respectively. 
Prior to the diffraction experiments the sample was dried for 96 hours at 1173\ K under argon flow and loaded in a controlled helium atmosphere using a glove box.
The combination of X-ray and neutron diffraction data proved to be crucial as the former is sensitive to the heavy cations, while the neutron scattering lengths for W (4.86\ fm), La (8.24\ fm) and O (5.80\ fm) are comparable. Rietveld refinement of crystal structure models against the experimental data was performed using GSAS with the EXPGUI graphical user interface.\cite{larson} High energy powder X-ray diffraction data were collected using ID15B also at the ESRF. A wavelength of 0.1422\ \AA \ was used and the scattered X-rays were detected by a Mar345 image plate. The pair distribution function was calculated using in-house software (iPDF) developed by one of the authors (SAJK), which runs as a graphical user interface within Igor Pro. Briefly, data were corrected for background, Compton scattering, and the atomic form factor. The Compton shift, detector efficiency and incoherent fluorescence were also taken into account, before Fourier transformation according to:
\begin{equation}
G(r)=\frac{2}{\pi}\int_{0}^{\infty}Q[S(Q)-1]sin(Qr)dQ
 \end{equation}
Here Q.[S(Q)-1] represents the properly corrected and normalized intermediate structure factor and the \textit{r}-grid used in real space had a spacing of 0.01\ \AA.
Models were fitted to the PDF data using the PDFgui package.\cite{farrow} Model PDFs were calculated using: 
\begin{equation}
G(r)=\frac{1}{Nr}\sum_{i}\sum_{j\neq i}\left[ \frac{b_{i}b_{j}}{\langle b \rangle^{2}}\delta(r-r_{ij})\right]-4\pi r \rho_{0}
 \end{equation}
The indices \textit{i} and \textit{j} run over all atoms in the sample. The scattering powers of the different atoms are \textit{b$_{i}$} and \textit{b$_{j}$}, and \textit{r} represents the radial distance in real space. PDFgui uses the so-called small box approximation, which implies that the first summation above only runs over the atoms within one unit cell as defined by the average crystallographic structure. This approximation makes data modeling tractable out to relatively large distances in real space. \\
\textit{EXAFS experiments}: Extended X-ray absorption fine structure spectroscopy (EXAFS) was performed on beamline KMC-2 at BESSY II, Berlin, on the W L$_{3}$ absorption edge (E\ =\ 10207\ eV) at room temperature and on beamline \textit{X} at Hasylab, Hamburg, on the La K absorption edge (E\ =\ 38925\ eV) at 10\ K. The measurements were performed in transmission mode and the data was processed and analyzed using the IFEFFIT code and the corresponding user interfaces ATHENA and ARTEMIS.\cite{ravel} After background subtraction, the normalized and k$^{3}$ \ weighted spectra were Fourier transformed over a photoelectron wave number range of k\ =\ 3.4\ -\ 12.6\ \AA$^{-1}$ for the W L$_{3}$-edge and k\ =\ 3.66\ -\ 16.46\ \AA$^{-1}$  for the La K-edge. The theoretical EXAFS signal was calculated by performing ab-initio calculations using the code FEFF8.2.\cite{ankudinov} The model was least squares fitted to the data in q-space with the following fitting parameters: a single amplitude reduction factor \textit{S$_{0}$$^{2}$}  and an overall energy parameter \textit{$\Delta$E$_{0}$} for each dataset, fractional changes in bond length $\alpha$ and a mean squared displacement parameter $\sigma$$^{2}$ for each coordination shell. \\
  \section{\label{sec:level1}RESULTS}
 
 \subsection{\label{sec:level2}Determination of the average crystal structure}
 \noindent\textit{Initial steps}:\\
The results of our neutron activation analysis showed that our sample had a composition of La$_{5.4(2)}$WO$_{11.1(2)}$. Analysis of the ICP-OES measurements gave a similar result. Our preliminary laboratory X-ray powder diffraction data showed that a doubled fluorite cell was found, with systematic absences consistent with face centred space groups.
 \begin{figure}[tb!]
\begin{center}
\includegraphics[scale=.31]{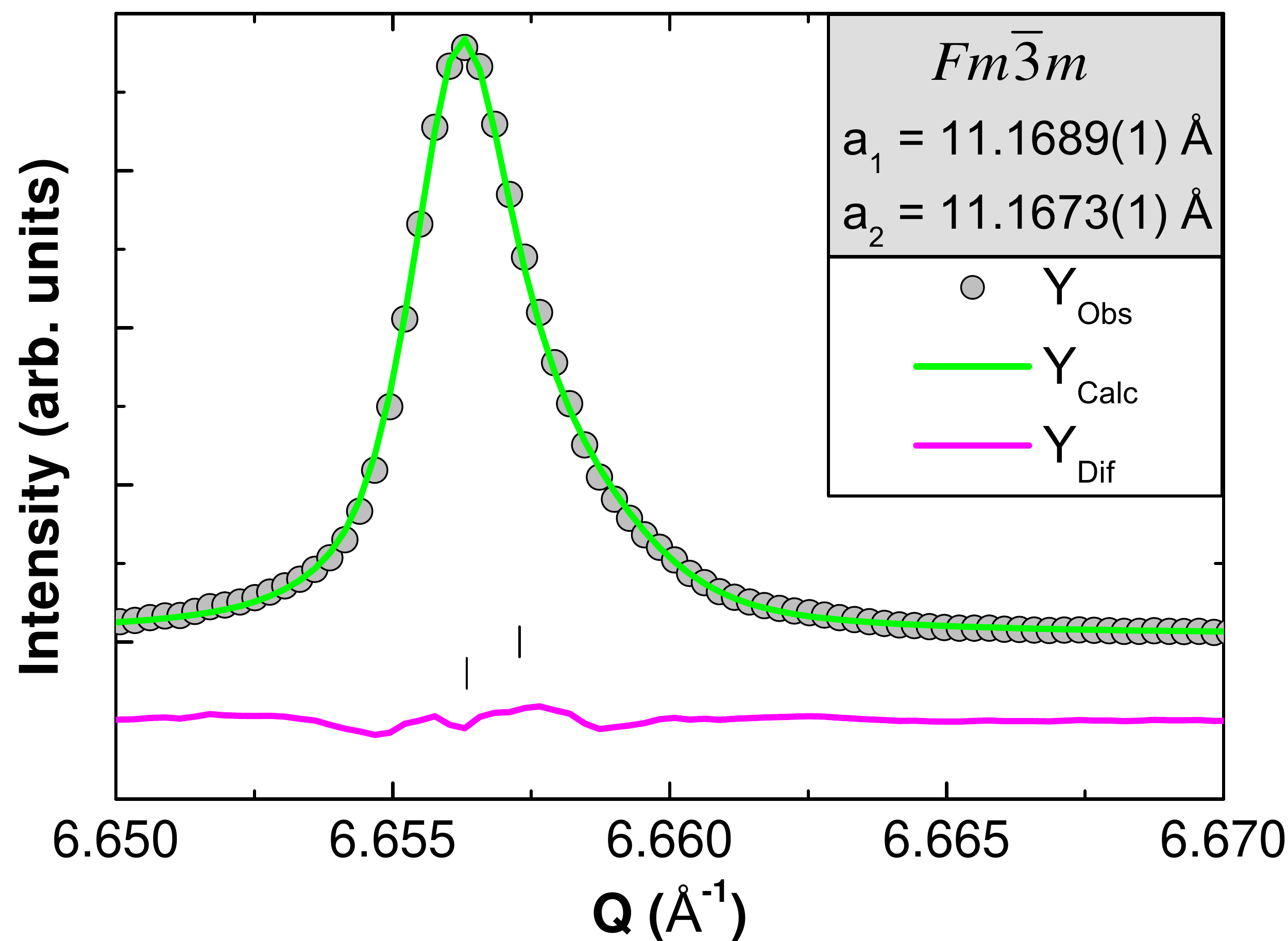}
\caption{(color online) Observed, calculated and difference profiles for the Rietveld fit to the synchrotron X-ray diffraction profile of \la, highlighting the region of the (10,6,2) reflection. The peak asymmetry is well fitted by an anisotropic broadening model as discussed in the text.}
\label{Fig1}
\end{center}
\end{figure}
 \begin{figure}[tb!]
\begin{center}
\includegraphics[scale=1]{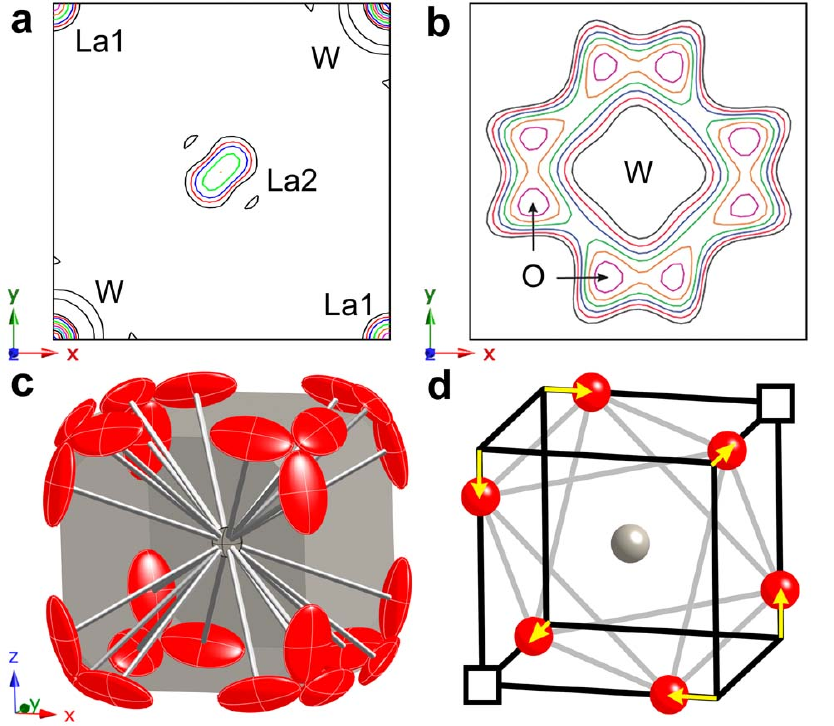}
\caption{(color online) (a) Electron density map calculated from HRSXRD data collected at 10 K. The map size is one half of the unit cell and the centre is at \textit{x}\ =\ \textit{y}\ =\ 0.25 and \textit{z}\ =\ 0. Contour lines with the same electron density are drawn at 1, 5, 10, 20, 30, 40, 50, 60, 70 and 80\ \AA$^{-3}$. (b) Fourier difference maps calculated from neutron powder diffraction data collected at 573\ K displaying the oxygen site splitting around W. The map size is 4\ x\ 4\ \AA$^{2}$ and the centre is at \textit{x}\ =\ \textit{y}\ =\ 0 and \textit{z}\ =\ 0.125. Contour lines are drawn at 0.05, 0.10, 0.15, 0.20, 0.25, and 0.30\ fm\AA$^{-3}$. (b) Resulting model for the oxygen site splitting around W (gray). The oxygen split site (red) is shown with its anisotropic atomic displacement parameters (ADPs) at 573\ K. (d) [111] oxygen vacancy pairs around W and the resulting oxygen displacements.}
\label{Fig1}
\end{center}
\end{figure}
\begin{figure}[tb!]
\begin{center}
\includegraphics[scale=1]{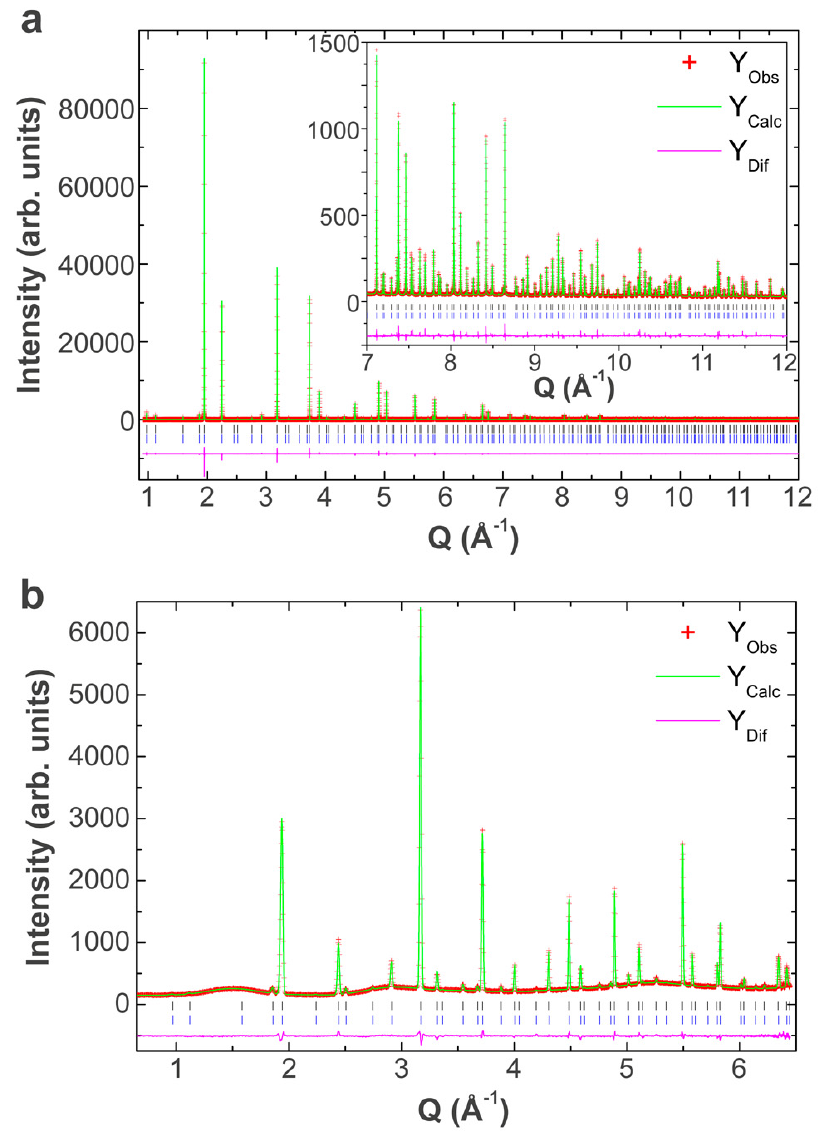}
\caption{(color online) Observed, calculated and difference plots for the Rietveld fits to the high-resolution X-ray powder diffraction data (a) and the neutron powder diffraction data (b). The refined parameters are given in Table I. Note that the sample was contained in a silica ampoule for the neutron diffraction measurements. The background in Fig. 3b is therefore not directly comparable to that shown in Fig. 5.}
\label{Fig1}
\end{center}
\end{figure}
Due to the limited Q-range data and medium resolution of the laboratory X-ray diffraction data, the structure solution was performed using the high-resolution synchrotron powder diffraction data (HRSXRD). Trial Le Bail intensity extractions using the doubled fluorite unit cell resulted in poor fit-statistics ($_{w}$R$_{p}$=12.14\ \% and $\chi^{2}$=6.50) and close examination of the diffraction profile showed that this was caused by anisotropic peak broadening, which affected all reflections equally (Fig. 1). As various models have been proposed in the literature\cite{mag:2009,mag:2012,mag:2013}, we performed an exhaustive search of distorted cells using Le Bail intensity extractions. This was compared with a minimal anisotropic broadening model which consisted of two cubic unit cells with an approximately 0.02\%  difference in lattice parameter (\textit{a}$_{1}$=11.1689(1)\ \AA, \textit{a}$_{2}$=11.1673(1)\ \AA \ at 10 K). None of the distorted cells tested, including tetragonal (\textit{I4/mmm}, no.\ 139), orthorhombic (\textit{Immm}, no.\ 71) or monoclinic (\textit{I112}, no.\ 5), gave better fit-statistics than the phase separation model. This result is significant because the lower symmetry cells have more variables and so a higher degree of freedom in the intensity extractions. Note that ID31 is perhaps the highest resolution powder diffractometer currently available, and that this tiny effect is invisible with other instrumentation. We note that the clear peak splittings observed in the ID31 data of Magraso \textit{et al} are almost certainly the result of measuring a phase separated sample, rather than a tetragonal distortion as reported in ref. [20].\\

\noindent\textit{Cation disorder}:\\
During the structure solution and refinement, we initially used two \textit{Fm$\bar{3}$m} cubic phases differing only in lattice parameter. Internal parameters were constrained to be the same for both phases. Electron density maps identified three heavy atom positions, corresponding to the Wyckoff sites 4\textit{a}\ (0,0,0), 4\textit{b}\ ($\frac{1}{2}$,$\frac{1}{2}$,$\frac{1}{2}$) and 24\textit{d}\ (0,$\frac{1}{4}$,$\frac{1}{4}$). These sites were attributed to the cations and, in the first cycles of Rietveld refinement, assumed to be statistically occupied with a La/W ratio of 5.4 as determined by NAA. We initially assumed fluorite-type oxygen sites  (32\textit{f}\ (\textit{x},\textit{x},\textit{x}) position, x$_{O(1)}$\ $\approx$\ 0.125 and x$_{O(2)}$\ $\approx$ \ 0.375). After these initial steps, we then exploited the differing contrasts of the X-ray and neutron data sets to order the cations such that the 4\textit{a} site with the highest electron density was occupied by W, with the 4\textit{b} site occupied by La and the 24\textit{d} site occupied with 98\% La and 2\% W. For the 24\textit{d} site, the refined atomic displacement parameters (ADPs) were large and anisotropic. Our electron density maps show that this site is actually split  (Fig. 2a) onto the 48\textit{h} (0,\textit{y},\textit{y}) site with 50\% occupancy. This disorder gave a much-improved refinement and realistic ADPs. These $\sim$0.22\ \AA \ displacements are static and localized as they were identical in the 10\ K and 295\ K data-sets.\\

\noindent\textit{Anion disorder}:\\
Finally, we were able to develop a disordered model for the oxygen sites coordinating the W 4\textit{a} site. Difference Fourier maps were calculated from the neutron powder diffraction data (Fig.\ 2b). These showed that oxygen occupies the 96\textit{k} site, with a total of 24 possible nearest neighbours arranged around the corners of a cube (Fig.\ 2c). The refined occupancy was found to be 25\%, which is equivalent to an average coordination of 6.0(1), as expected for octahedral W$^{6+}$. The oxygen displacement parameters were found to be anisotropic, and elongated in the direction transverse to the W-O bonds.
\begin{figure}[tb!]
\begin{center}
\includegraphics[scale=1]{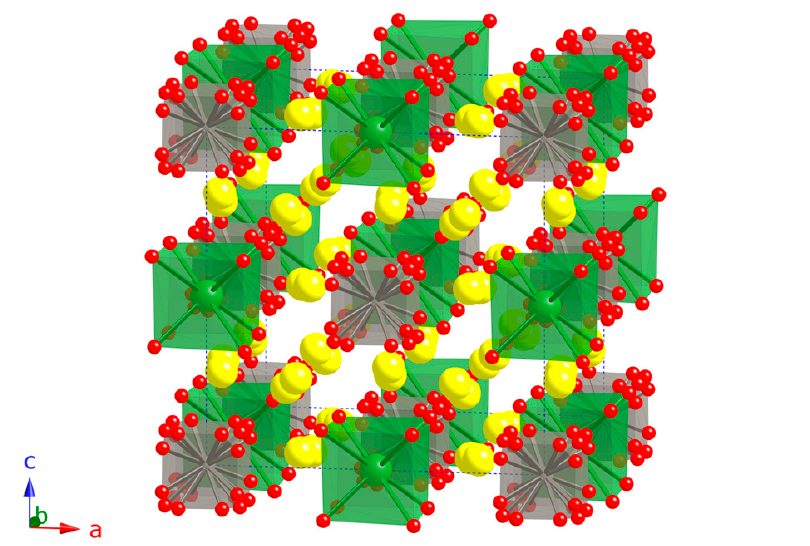}
\caption{(color online) The average crystal structure refined for \la \ is shown. Red spheres represent oxygen atoms, gray cubes W polyhedra with their partially occupied oxygen sites, green cubes the La(1) sites and yellow spheres the half occupied La(2) split sites. About one out of 48 La(2) positions is occupied by W, yielding the correct composition of \la.}
\label{Fig1}
\end{center}
\end{figure}
The local coordination can be explained by consideration of the defect chemistry. Assuming two vacancies per site, and that it is energetically favourable to locate them as far apart as possible, gives the picture shown in Fig.\ 2d. The vacancy pair is located on one [111] axis of the cube. The nearest-neighbour anions can then relax toward the defect in the [100] direction by 0.5\ \AA, forming a nearly regular octahedra. In combination with the cation disorder described above, this model then gave an excellent fit to both the synchrotron X-ray and neutron powder diffraction profiles (Figures\ 3a and 3b). The refined atomic coordinates, displacement parameters and residuals from the fit to the former are given in Table\ I.
The final model for the average crystal structure of \la \ is shown in Fig. 4. The structure is derived from a fluorite structure with doubled lattice parameter due to cation ordering. The tungsten positions form an \textit{fcc} lattice and nearest W neighbors are linked by the split La(2)/W(2) site, which has an average of 7 fold oxygen coordination. In the structural model developed here, one W is substituted onto the La(2)/W(2) site 48\textit{h} to stabilize the \la \ phase. This is in complete contrast to Ln$_{6}$WO$_{12}$ materials containing rare earth atoms with small ionic radii, which show complete ordering of cations and oxygen vacancies.\cite{diot}
\begin{table}
\caption{\label{tab:table2}Refined atomic coordinates for \la \ at 295 K from synchrotron X-ray powder diffraction in space group \textit{Fm$\bar{3}$m}. The site occupancies are W(1) = 1, La(1) = 1, La(2)/W(2) = 0.48/0.02, O(1) = 0.25, O(2) = 0.9825. The refined average lattice parameter is $a$ = 11.1844(1) \AA \ and the residuals are $_{w}$R$_{p}$\ =\ 7.36\% , $\chi^{2}$\ =\ 2.34 and R$_{F^{2}}$\ =\ 3.40\%. }
\begin{ruledtabular}
\begin{tabular}{ccccc}
 Atom&$x$&$y$&$z$&uiso (\AA$^{2}$x100)\\
 \hline
W(1)&0&0&0&0.787(6)\\
La(1)&0.5&0.5&0.5&1.71(1)\\
La(2)/W(2)&0&0.23622(1)&0.23622(1)&1.238(5)\\
O(1)&0.1108(2)&0.1108(2)&0.0663(3)&1.7(1)\\
O(2)&0.3662(1)&0.3662(1)&0.3662(1)&1.92(4)\\
      \end{tabular}
\end{ruledtabular}
\end{table}
  \subsection{\label{sec:level2}Local structure determination}
\noindent \textit{Development of the local model}:\\
The model described in the previous section gives an excellent fit to the Bragg reflections of \la. However, we detected strong diffuse scattering in the background of the neutron diffraction data sets (Fig. 5) which hints of local order.  This oscillatory contribution is invisible to X-rays and hence likely originates from oxygen-oxygen correlations. To explain this signal, we returned to the defect model and considered how octahedral order might propagate through the lattice. As shown in Fig. 6a, the orientation of a WO$_{6}$ octahedra influences that of its neighbours in a manner analogous to the way a magnetic exchange interaction links the orientation of localised spins in magnetic insulators. As described previously and shown in Fig. 2d and Fig. 6a, removing the oxygen atoms on one diagonal of a W polyhedron results in two oxygen vacancies and a relaxation of the remaining oxygen atoms towards the vacancies to form a regular octahedron. The oxygen vacancies also reduce the coordination number of the neighbouring La(2) site.
 \begin{figure}[tb!]
\begin{center}
\includegraphics[scale=1]{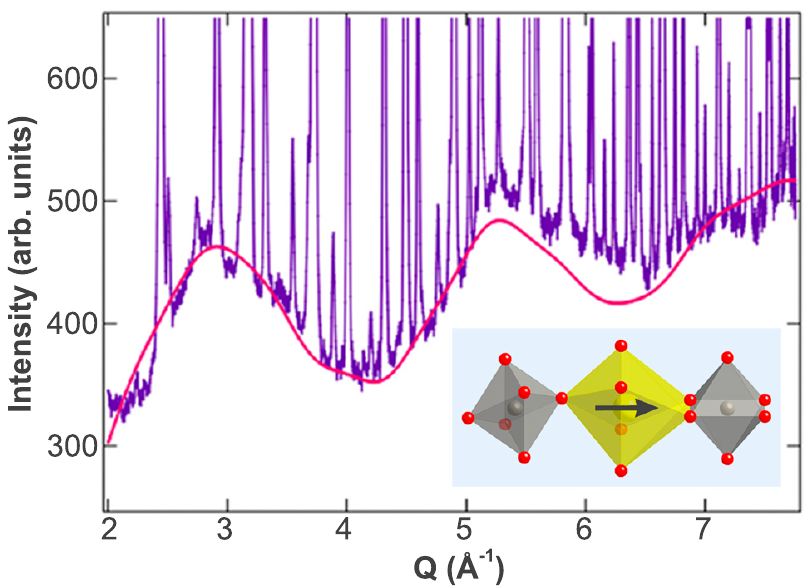}
\caption{(color online) Diffuse scattering of neutrons is observed at room temperature for dry samples of \la. The red line is a calculation of scattering from local order over the pair of sites shown in the inset using equation (3) and a background $\propto$Q$^{2}$ to account for phonon scattering. Note that the sample was contained in a vanadium can for this measurement. This sample container only contributes a structureless incoherent background to this data set.} 
\label{fig2}
\end{center}
\end{figure}
As six-fold coordination is energetically disfavoured for La, this site then relaxes away from the defect toward the nearest neighbour WO$_{6}$ octahedra. This model therefore explains why the La(2) site is split along the line connecting two W sites. The possible orientations of the next WO$_{6}$ octahedra are then reduced, as shown in Fig.\ 6a.
 \begin{figure}[tb!]
\begin{center}
\includegraphics[scale=1]{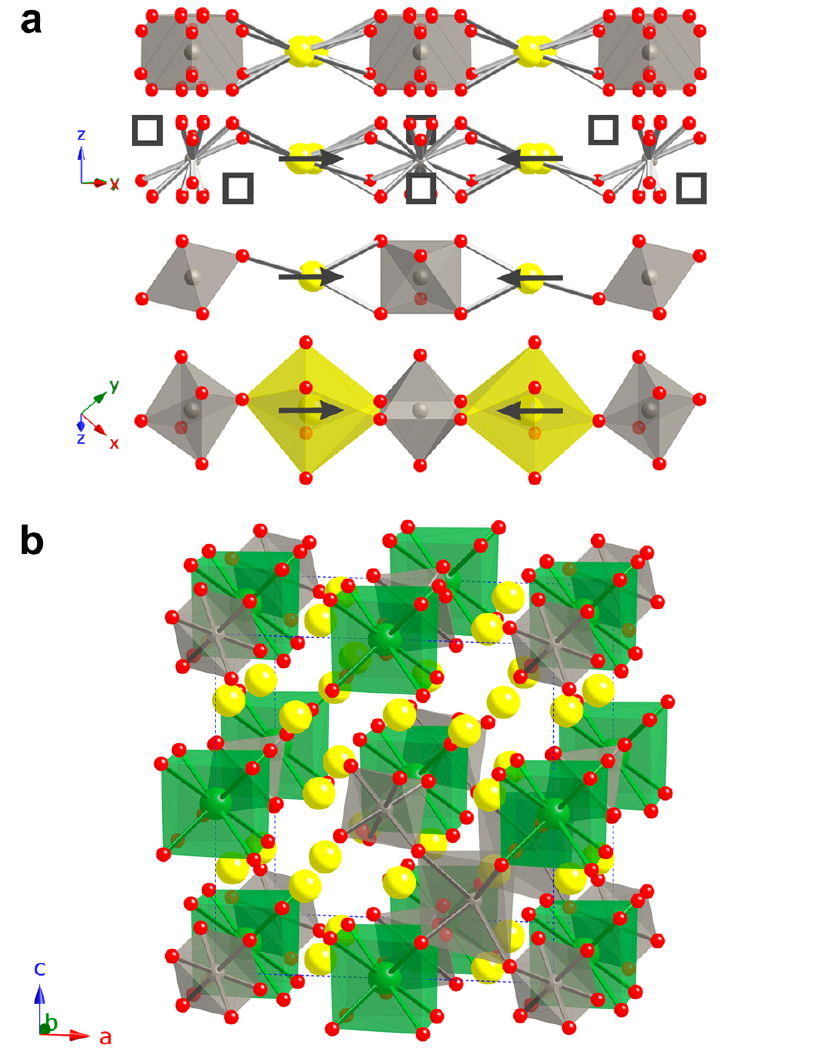}
\caption{(color online) (a) Development of a possible local configuration (bottom) from the average crystal structure (top) with La and O split sites. Fixing the orientation of a single WO$_{6}$ octahedron propagates order to nearest neighbors. To preserve a La coordination of seven, La must share one oxygen defect, resulting in a displacement of La as shown by arrows. (b) Local crystal structure for \la \ in space group \textit{Pa$\bar{3}$} (no. 205). Red spheres represent oxygen atoms, gray W octahedra, green cubes the La(1) sites and the yellow spheres La(2) sites. One out of 24 La(2) positions is occupied by W, ensuring the correct composition \la.}
\label{Fig1}
\end{center}
\end{figure}

The importance of this novel 'chemical exchange interaction' is underlined by the simulation shown in Fig.\ 5. Here we calculated the powder averaged scattering S(Q) of the dimer unit shown in Fig.\ 6a using the Debye equation, which assumes that the static approximation holds:
\begin{equation}
S(Q)\propto\frac{1}{N}\sum_{i}\sum_{j}b_{i}b_{j}\frac{sin(Q.r_{ij})}{Q.r_{ij}}
 \end{equation}
where $b_{i}$ and $b_{j}$ are the neutron scattering lengths, Q is the scattering vector, $r_{ij}$ is the interatomic distance between two atoms and $N$ is the number of atoms. This model reproduces the position of the maxima in the background and suggests that correlations between the orientation of WO$_{6}$ octahedra extend beyond at least one shell.\\

\noindent\textit{Pair distribution function analysis}:\\
As the above modelling of the diffuse scattering is only qualitative, we compared the average and local structures to the X-ray pair distribution function (PDF) of \la \ at room temperature. While the X-ray PDF is not especially sensitive to the position of the oxygen atoms, it is highly sensitive to the displacements of the La(2) sites as we describe in more detail below. The so-called intermediate structure factor, Q.[S(Q)-1], of \la \ is shown in Fig. 7. The signal is seen to be damped to near zero above Q = 15 \AA$^{-1}$ as a result of disorder. We truncated this data set at Q = 24 \AA$^{-1}$, and performed the Fourier transform described in the experimental methods section. This yields the PDF shown in the bottom panel of Fig. 7. In order to determine the length scale of the local order, we performed refinements of two structures against this data set in the range 1.5 $<$ \textit{r} $<$ 12 \AA. Models were fitted in the small box approximation described in the experimental section and the calculated PDF was convoluted with the Fourier transform of the box function representing the measured Q-range. Isotropic thermal displacement parameters were used and we applied a sharpening correction for correlated motion in the first coordination sphere. The observed damping of the PDF in real space (which is due to the instrumental resolution) was accounted for by measuring a standard (NIST CeO$_{2}$). \\
The first structure we refined against the data was the average \textit{Fm$\bar{3}$m} structure determined above. 
\begin{figure}[tb!]
\begin{center}
\includegraphics[scale=.55]{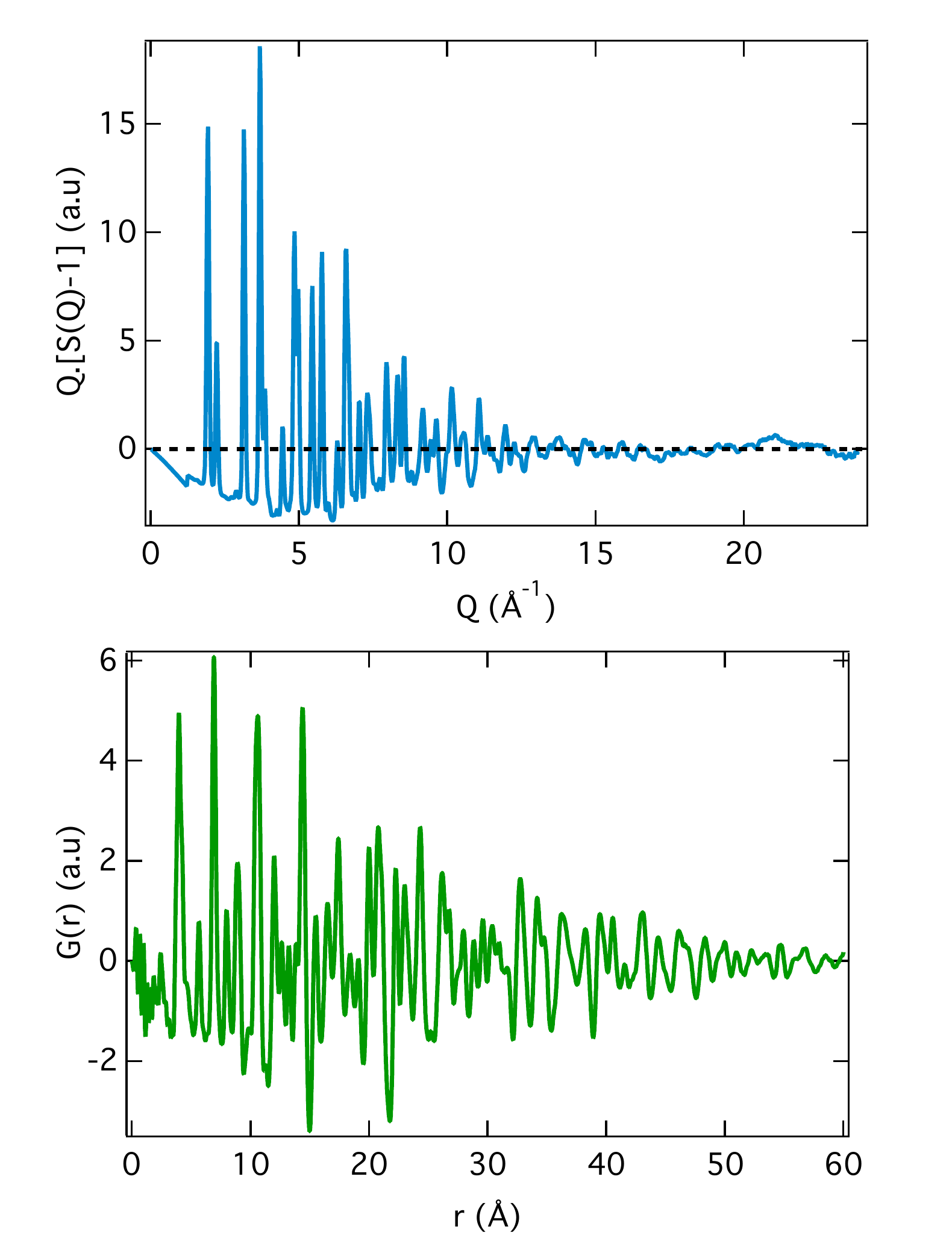}
\caption{(color online) (top) Corrected intermediate structure factor, Q.[S(Q)-1], for \la, calculated from the high energy powder diffraction data as described in the text. (bottom) Pair distribution function for \la \ calculated in the range 0 $<$ \ $r$ \ $<$ \ 60 \AA \ by Fourier transform of the data shown in the top panel. The $r$-grid used was 0.01 \AA. } 
\label{fig2}
\end{center}
\end{figure}
\begin{figure}[tb!]
\begin{center}
\includegraphics[scale=0.25]{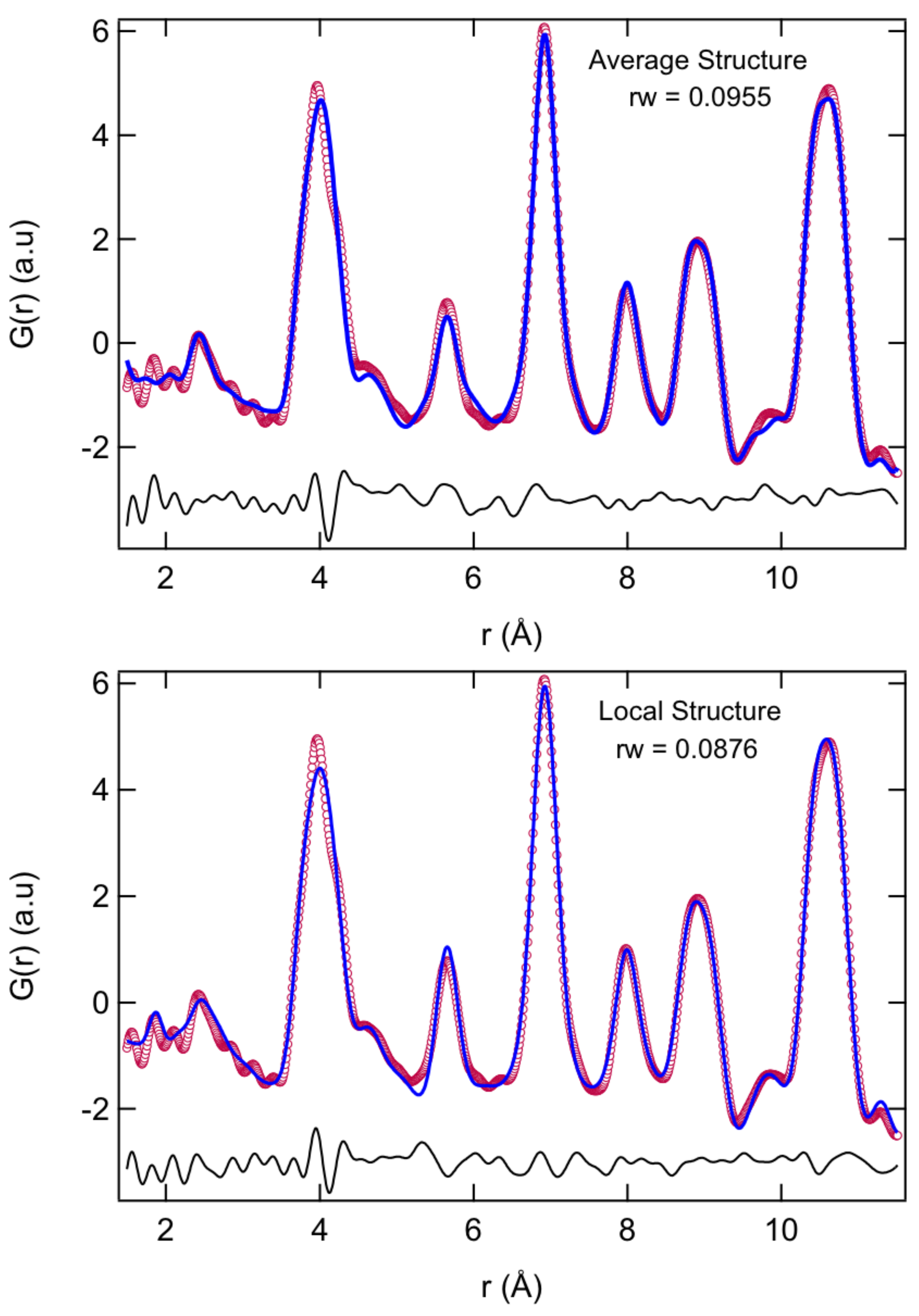}
\caption{(color online) Observed, calculated and difference plots for the real space fits to the X-ray pair distribution function of \la \ at room temperature. The fit to the \textit{Fm$\bar{3}$m}  structure is shown in the top panel, and the fit to the \textit{Pa$\bar{3}$} structure is shown in the bottom panel.} 
\label{fig2}
\end{center}
\end{figure}
\begin{figure}[tb!]
\begin{center}
\includegraphics[scale=.23]{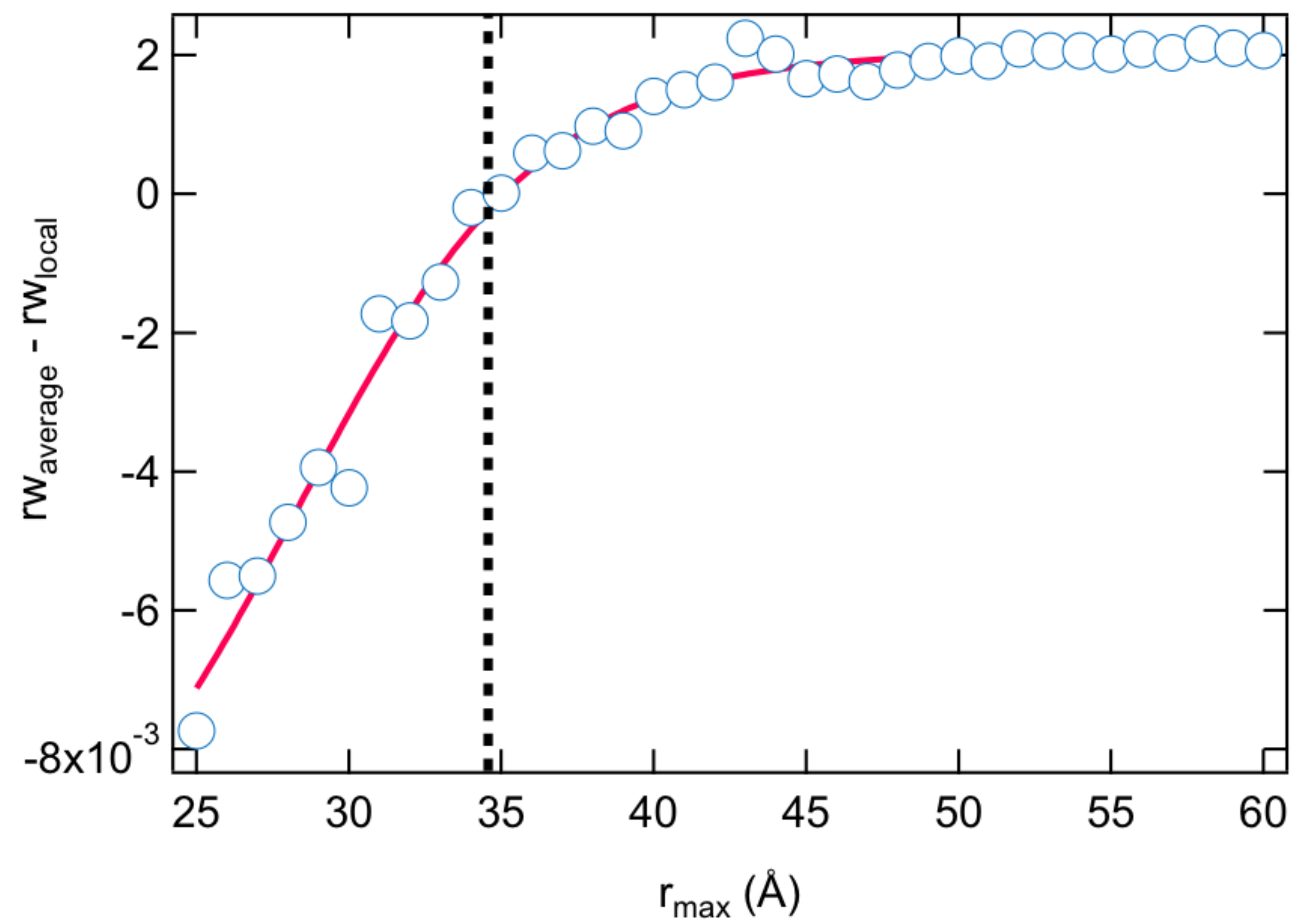}
\caption{(color online) The local structure model describes the pair distribution function of \la \ better that the average structure model up to distances of 35 \AA. The plot shows the difference between the residuals of the two models as a function of the maximum \textit{r} value fitted.} 
\label{fig2}
\end{center}
\end{figure}
This fitted well (rw = 0.0955), but with some misfits in the first coordination shell (see the peak at $\sim$2 \AA \ in Fig. 8a). We therefore generated a new structure in \textit{Pa$\bar{3}$}, which consisted of ordered WO$_{6}$ octahedra and lanthanum displacements, as shown in Fig. 6b. Despite the fact that this structure does not model any of the average  structure displacements of La or O atoms, it not only provides a better fit (rw = 0.0876), but also gives a chemically reasonable W-O bond length of 1.87 \AA. The corresponding sharp peak in the PDF (Fig. 8b) reflects the strongly covalent nature of these bonds and validates the earlier assumptions about octahedral coordination. The local coordination units in \la \ can therefore be regarded as rigid bodies, rather than the highly distorted and disordered units inferred from the average structure.The refined coordinates from this model are shown in Table II. \\
\begin{table}
\caption{\label{tab:table2}Refined atomic coordinates for \la \ at room temperature from synchrotron X-ray pair distribution function analysis in space group \textit{Pa$\bar{3}$}. The lattice parameter, $a$ = 11.299 \AA. Site occupancies are as before.}
\begin{ruledtabular}
\begin{tabular}{cccc}
 Atom&$x$&$y$&$z$\\
 \hline
W(1)&0&0&0\\
La(1)&0.5&0.5&0.5\\
La(2)/W(2)&0.2366&0.7633&0.999\\
O(1)&0.9209&0.8699&0.0654\\
O(2)&0.6222&0.6222&0.6222\\
O(3)&0.6567&0.3606&0.3742\\
      \end{tabular}
\end{ruledtabular}
\end{table}
While both models give good fits to the PDF data up to $\sim$10 \AA, we found that the fit of the average model was better at longer distances. This effect arises because different \textit{r}-ranges of the pair distribution function are sensitive to order on different length scales. We therefore performed a series of structure refinements where the fitted maximum ($r$-max) increased in a stepwise fashion. We could then compare the residual values for the two models and estimate the length scale of the local order in \la. The results of this analysis are shown in Fig.\ 9. The difference rw$_{average}$-rw$_{local}$ is negative until the $r$-max exceeds 35 \AA. When data at longer distances is included, the average structure determined by Bragg diffraction becomes more favourable. It should be noted that the refined \textit{Pa$\bar{3}$} structure completely failed to describe the data sets shown in Fig.\ 3 when we attempted Rietveld refinements. This confirms that the length scale for the order in \la \ is restricted to small domains of the order of 3.5 nm, which is much below the coherence length required to yield sharp Bragg reflections. We note that our analysis provides no information on the shape or orientation of these domains as our model assumes uncorrelated spherical regions.\\

\noindent\textit{Extended X-ray absorption fine structure spectroscopy (EXAFS)}:
\begin{figure*}[tb!]
\begin{center}
\includegraphics[scale=1]{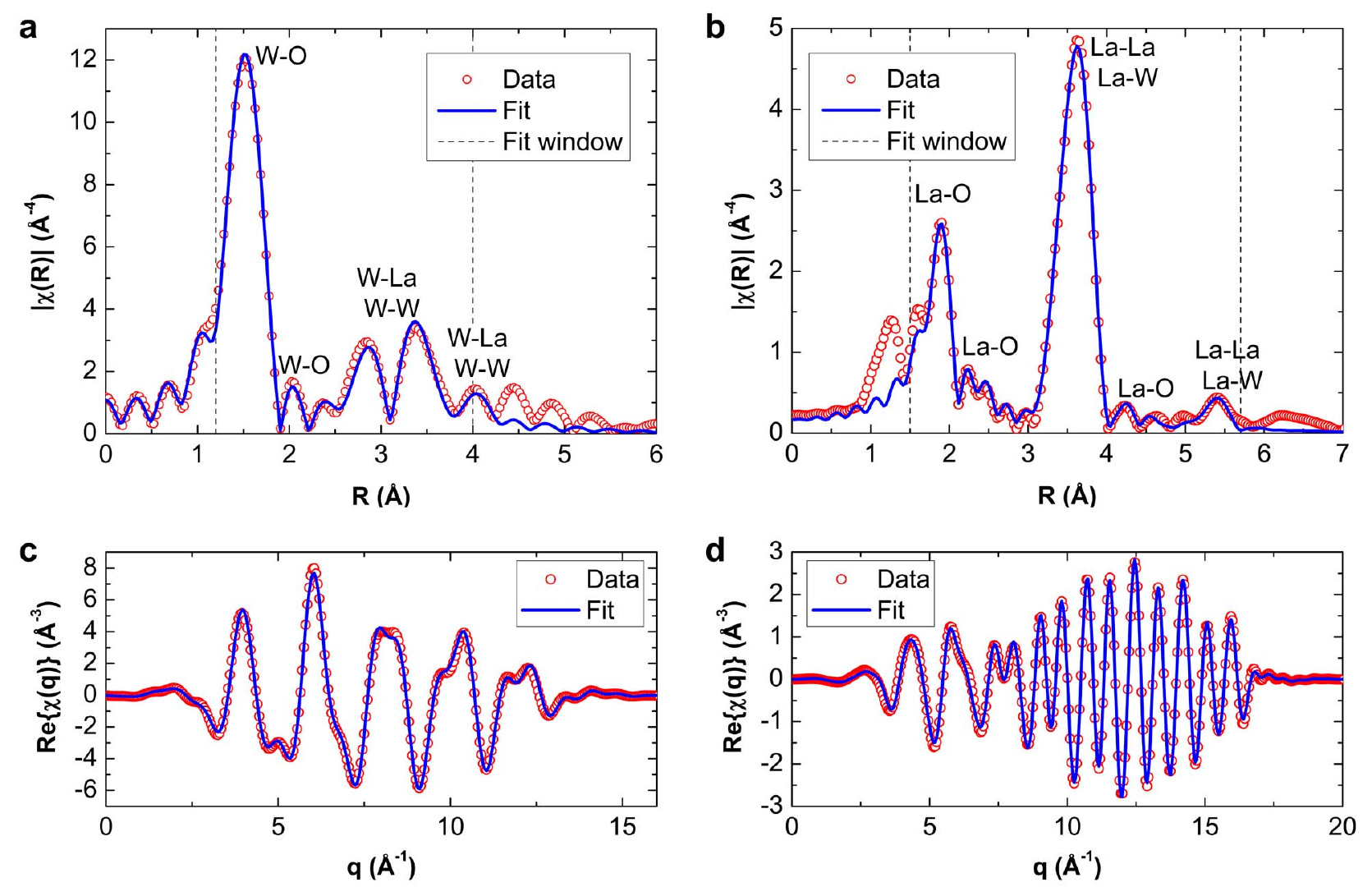}
\caption{(color online) Fourier transformed $k^{3}$ weighted EXAFS data as a function of the radial distance (uncorrected for phase shifts) together with the fit and the fitting window are shown for (a) W L$_{3}$-edge at 300 K and (b) La K-edge at 10 K. The fits were performed in \textit{q}-space (c) and (d) for the local crystal structure in \textit{Pa$\bar{3}$} (see Figure 6b).}
\label{fig2}
\end{center}
\end{figure*}
In order to provide an element specific confirmation of our proposed local structure, we turn now to the results of our EXAFS experiments on the W L$_{3}$ and La K-edges. The former also provided confirmation of the small amount of anti-site disorder on the La(2) sites, which is close to the minimum measurable by diffraction techniques. Both of the data sets provided information out to $\sim$6 \AA \ as shown in Fig.\ 10. Fitting of the models to the data was perform in \textit{k}-space. As a starting model we used the \textit{Pa$\bar{3}$} structure refined against the pair-distribution function data above. For both edges, the shells were fitted in \textit{k}-space one after another, and as a final step all shells were fitted together.\\
At the W L$_{3}$-edge, we compared two different variations on the \textit{Pa$\bar{3}$} structure. The first model had a nominal La/W ratio of 7 (4 W per unit cell) contains only one W site at 4\textit{a}. This refinement gave a rather poor \textit{R}-factor of 3\%. However, a dramatic improvement was observed when anti site disorder was considered in the second model. We used a nominal La/W ratio of 5.4 (5 W per unit cell) contains two W sites. Four W were placed on the 4\textit{a} site and one additional W was substituted onto the La(2) 24\textit{d} site. This fit gave a dramatically improved \textit{R}-factor of 0.3\%. The  Fourier transformed and \textit{k$^{3}$} weighted EXAFS data, together with the best fit line from this model are shown in Figs.\ 10a and 10c. The second W site is manifested by the W-La peak at 2.8 \AA, as well as the split W-W distance at 3.5 and 4 \AA. Note that these data are not phase corrected, and the true bond lengths are given in Table III. \\
For refinements against the La K-edge EXAFS data, we used the same structural model as developed above. This model contains 27 La atoms on two La sites, four La on site 4\textit{b} and 23 La on site 24\textit{d}. The refinement was performed with all scattering paths for the two La sites up to 5.6  \AA \ (36 paths for two La sites). Once again, we achieved excellent agreement between the data and the model calculation, with an R-factor of 0.31 \%. The Fourier transformed and k$^{3}$ weighted EXAFS data, together with the best fit line from this model are shown in Figs. 10b and 10d. The bond lengths extracted from this refinement are also given in Table III. In the final structure, the La(1) site has a highly regular environment of 8 oxygen ions arranged on the corners of a cube and 12 La neighbours. The surrounding of the La(2) site however shows a high degree of disorder. La(2) is sevenfold coordinated by oxygen and the bond distances are between 2.37 \AA \ and 2.67 \AA. The second coordination shell comprises 12 metal neighbours and is split into seven sub shells between 3.66 \AA \ and 4.22 \AA. Due to the disorder of the second shell, the metal-metal peak shown in in Fig.\ 10b at 3.5 \AA, is very broad and the signal is strongly damped at higher radial distances. Nevertheless, as our split site model describes the local structure well, the Debye-Waller factors remain small, showing no contribution from un-modelled static disorder.\\
A comparison of the results of fitting the W L$_{3}$ and La K-edges shows how the anion sub-lattice relaxes around the W anti-site defects. For example, we find a contraction of the metal-oxygen distance in the first coordination shell when La$^{3+}$ is replaced by the smaller and more highly charged W$^{6+}$ cations. Indeed the shortest bond from the 24\textit{d} site is reduced from 2.373(7) to 1.839(16) \AA. The latter value is in excellent agreement with the W-O distance extracted for the fully occupied 4\textit{a} site (1.908(2) \AA), as well as that found by our PDF experiments (1.87 \AA). Our EXAFS results therefore fully support the local structure we developed for \la \ by diffraction measurements, as well as confirming the presence of the $\sim$4\% anti-site disorder on the La(2) site.
\begin{table*}
\caption{\label{tab:table2}Refined bond distances and coordination numbers from fitting of the La K-edge (10 K) and W L$_{3}$-edge (300 K) EXAFS signal of \la. The \textit{Pa$\bar{3}$} local structure model was used as a starting point.}
\begin{ruledtabular}
\begin{tabular}{cccccc|cccccc}
 Atom&1$^{st}$ shell La-O&CN&$2^{nd}$ shell La-M&CN&M&Atom&1$^{st}$ shell W-O&CN&2$^{nd}$ shell W-M&CN&M\\
 \hline
La(1) on 4\textit{b}&2.533(9)&8&3.92(3)&12&La&W(1) on 4\textit{a}&1.908(2)&6&3.687(5)&6&La\\
&&&&&&&&&4.136(7)&6&La\\
La(2) on 24\textit{d}&2.373(7)&2&3.656(7)&1&W&W(2) on 24\textit{d}&1.839(16)&2&3.199(13)&1&W\\
&2.414(7)&2&3.759(12)&2&La&&1.871(16)&2&3.199(13)&2&La\\
&2.512(13)&1&3.953(13)&2&La&&2.25(4)&1&3.363(14)&2&La\\
&2.642(14)&1&3.963(7)&2&La&&2.36(5)&1&3.391(14)&2&La\\
&2.670(14)&1&3.978(7)&2&La&&2.39(5)&1&3.413(15)&2&La\\
&&&4.083(8)&1&W&&&&3.583(15)&1&W\\
&&&4.220(8)&2&La&&&&3.621(15)&2&La\\
\end{tabular}
\end{ruledtabular}
\end{table*}
\section{\label{sec:level1}DISCUSSION}
The above experiments have identified a complex hierarchy of structure, which spans multiple length scales in \la. In what follows, we briefly review our average structure in comparison to earlier work, and discuss the local structure in more detail. \\
The lattice symmetry of \la \ has previously been called into question based on very high resolution synchrotron x-ray powder diffraction experiments performed at the ESRF.\cite{mag:2012} However, our results show that \la \ is actually cubic, but with slightly asymmetric peak broadening that implies a tiny ($\sim$0.02 \%) lattice parameter distribution. We emphasise again that this effect is masked entirely by the resolution of standard X-ray and neutron diffraction instrumentation. Having accounted for this effect, we were able to propose a new average structure model in space group \textit{Fm$\bar3$m}, which has two improvements on previously published work. Firstly, we identified positional disorder on parts of the metal sublattice, which has an appealing explanation in terms of the defect chemistry in \la. We find (Fig.\ 2a) that the La(2) site is split into two 50:50 occupied sites with displacement along the vectors joining neighbouring tungsten sites. When the possible location of oxygen vacancies in the lattice (Fig. 2 and Fig. 6) are accounted for, this finds a natural explanation, as the La sites relax away from the defects to maintain seven-fold coordination. As we comment on more fully below, this is naively expected to propagate order through the lattice at temperatures below the onset of oxygen ion conduction. The second improvement of our model on previous work is the quantification of anti-site disorder. The possibility of a small amount (ca. 5\%) of tungsten subsitution on the La(2) sites was suggested from DFT calculations. However, no direct evidence of this has yet been presented. Here we used a combination of x-ray and neutron diffraction, and exploited the differing contrasts of these techniques to get a consistent estimate of the anti-site disorder. When we refined the occupancy of the La(2) site against the X-ray data, it rose to above unity, consistent with a small amount of heavier W. Indeed the refined X-ray scattering power was equivalent to 53.32 e$^{-}$, where f$_{1}$(La$^{3+}$) = 52.79e$^{-}$ and f$_{1}$(W$^{6+}$) = 68.24e$^{-}$). For the refinement against the neutron diffraction data, we found an average scattering length of  8.12 fm, reduced from the value expected for La (\textit{b}$_{coh}$=8.24 fm), again consistent with substitution by W (\textit{b}$_{coh}$ = 4.86 fm). These measurements thus enabled us to calculate an anti-site disorder of 4.4(2)\%. In addition, out EXAFS measurements on the W L$_{3}$-edge, which are much more sensitive to this small structural perturbation, gave an additional confirmation. The unique element specificity of EXAFS means that extra sites are immediately apparent. Here we found an order of magnitude decrease in the fitting residuals of our structure refinement when we included the trace amount of disorder inferred from the diffraction measurements.\\
While the details described above complete the description of the average structure of \la, our results go much further on local length scales. We were able to obtain qualitative insight into the local structure by using simple chemical concepts. For example, from an electrostatic point of view, the approximately octahedral coordination of tungsten shown in Fig. 2d is by far the most likely as it minimises the repulsion between ligands. Similar effects have been inferred for other simple functional materials, such as scandium and yttrium stabilised zirconias.\cite{steele,irvine} This is shown by diffuse scattering observed in neutron powder diffraction experiments, which are sensitive to oxygen order. Here, a simple model of two linked octahedra accounts for the peak positions of this scattering (Fig. 5). This result suggests the intriguing possibility that order might propagate over a limited range, below the coherence length required to produce sharp Bragg peaks. Assuming that all octahedra have fixed orientations produces our \textit{Pa$\bar3$} \ structure (Fig. 6b). Once again, this insight from diffraction is strongly confirmed by our EXAFS data sets, which were also satisfactorily fitted by this model.\\
We completed our picture of the local structure using real-space refinements against our X-ray pair distribution function data. This provides a definitive model for the \textit{Pa$\bar3$} \ structure which will be of significant use for DFT investigations of the mechanism for mixed conduction. Furthermore, since we were able to Fourier transform our data out to at least 60 \AA, we could estimate the size of the coherent \textit{Pa$\bar3$} \ domains. This analysis yielded a local coherence length of ca. 3.5 nm. We anticipate that future PDF experiments performed at high temperatures, will enable determination of the link between this length scale and mixed conduction performance in \la.\\
Finally, the observation of local order of a limited length scale raises interesting questions. For example, why is this high entropy ground state favoured in \la? Furthermore, what factors prevent long-range order from establishing itself in this material? We conclude our discussion by briefly speculating about connections with a seemingly unrelated field, that of so-called frustrated materials.\cite{henley,bramwell} These are materials, in which pairwise interactions (for example an antiferromagnetic exchange interaction, J) compete due to the lattice connectivity. Long-range order is not established even at temperatures where \textit{k}$_{B}$ $<<$ J. The canonical example is water ice, where local rules enforcing the formation of bent water molecules create a massively degenerate ground state manifold. Here the picture is similar. In our local structure, the tungsten cations are placed on a non-bipartite \textit{fcc} lattice. When pairwise interactions are present (such as our local rule linking the orientation of WO$_{6}$ octahedra) this lattice is frustrated. We speculate that this also results in a hugely degenerate number of structures for \la, which results in the 'glassy' ground state we observe. Indeed, our average structure model is nothing more than the result of spatially averaging the orientation of all of the WO$_{6}$ octahedra in the sample. We note in passing, that much as the proton displacements in water ice map to an Ising spin model ('spin ice'),\cite{castro} the La displacements in \la \ map to a well-studied Ising model on the octahedral lattice.\cite{henley} This comparison may be relevant for future investigations of defect dynamics in \la. For example, the propagation of defects in frustrated models is often quite different to that in standard materials. For example, protons diffuse through ice,\cite{castro} rather than the hopping motion found in ceramic proton conductors.\cite{karlsson} We believe future investigations of the local interactions in \la \  and other mixed conductors are thus well-merited and speculate that so-called 'function through frustration' may be an appealing route to designing new mixed conductors in the future.

\section{\label{sec:level1}CONCLUSIONS}
In summary, we have reported a detailed study of both the average and the local structure of \la . We show that former has both positional and substitutional disorder of metal sites, as well as partially occupied anions sites. However, our diffuse scattering experiments show that on a local scale, rigid WO$_{6}$ octahedra are found and we propose a simple defect rule which links the orientation of neighbouring sites. Our local structure is confirmed by EXAFS experiments and will be of significant use in further modelling the mixed conduction of this class-leading material.

\section{\label{sec:level1}ACKNOWLEDGEMENTS}
We thank the Helmholtz Association for funding through the Helmholtz Alliance MEM-BRAIN (Initiative and Networking Fund) and the Spanish Government (grant ENE2011-24761). We thank D. Alber and G. Bukalis for NAA measurements and E. Suard and G. Nenert for assistance with neutron diffraction data collection. The European Synchrotron Radiation Facility, BESSY-II, Hasylab and the Institut Laue Langevin are thanked for the provision of beam time. We thank D.A. Tennant, T. Fennell, D.J.P. Morris and S.J.L. Billinge for useful discussions.

\end{document}